\documentclass[12pt]{iopart}
\usepackage{graphicx, caption}
\expandafter\let\csname equation*\endcsname=\relax 
\expandafter\let\csname endequation*\endcsname=\relax 
\usepackage{amsmath,amssymb,amsthm}
\usepackage{epsfig,subfigure}
\usepackage{color}
\usepackage{float}
\usepackage{hyperref}
\usepackage{times}
\usepackage{todonotes}

\hypersetup{
pdfauthor = {SB},
pdftitle = {Photo-emission signatures of coherence breakdown in Kondo alloys: dynamical mean-field theory approach \today},
}

\begin{document}

\title{Photo-emission signatures of coherence breakdown in Kondo alloys: dynamical mean-field theory approach}
\author{B Poudel$^1$, C Lacroix$^2$, G Zwicknagl$^3$ and S Burdin$^1$}
\address{$^1$ Universit\'e de Bordeaux, CNRS, LOMA, UMR 5798, 33400 Talence, France}
\address{$^2$ Institut N\'eel, CNRS and Universit\'e Grenoble-Alpes, Boite Postale 166, 38042 Grenoble Cedex 09, France}
\address{$^3$ Technical University Braunschweig, 0531 Braunschweig, Germany}



\date{\today}
\maketitle

\begin{abstract}
    We study the Kondo alloy model on a square lattice using dynamical mean-field theory (DMFT) for Kondo substitution and disorder effects, together with static mean-field approximations. We computed and analyzed photoemission properties as a function of electronic filling $n_c$, Kondo impurity concentration $x$, and strength of Kondo temperature $T_K$. We provide a complete description of the Angle Resolved Photoemission Spectroscopy (ARPES) signals expected in the paramagnetic Kondo phases. By analyzing the Fermi surface, we observe the Lifshitz-like transition predicted previously for strong $T_K$ at $x=n_c$ and we discuss the evolution of the dispersion from the dense coherent to the dilute Kondo regimes. At smaller $T_K$, we find that this transition marking the breakdown of coherence at $x=n_c$ becomes a crossover. However, we identify another transition at a smaller concentration $x^\star$ where the effective mass continuously vanishes. $x^\star$ separates the one-branch and the two-branches ARPES dispersions characterizing respectively dilute and dense Kondo paramagnetic regimes. The $x-T_K$ phase diagrams are also described, suggesting that the transition at $x^\star$ might be experimentally observable since magnetically ordered phases are stabilized at much lower $T_K$. Fermi surface reconstructions in antiferromagnetic and ferromagnetic phases are also discussed.  
\end{abstract}

\section{Introduction}

Kondo effect is a fundamental microscopic mechanism which is relevant for studying strongly correlated electron 
materials~\cite{hewson1997kondo,fulde2006strongly}. 
It is characterized by the low temperature formation of a highly entangled state between conduction electrons and quantum magnetic impurities. It was observed in rare-earth based compounds below the Kondo temperature $T_K$,
where Kondo impurities are realized by Kramers doublets of $4f$ or $5f$ orbitals, which may form local singlets with the spins of conduction electrons. In most cases, the low energy excitations of Kondo paramagnetic ground states can be described as Fermi liquids~\cite{nozieres1974fermi}. 

In concentrated Kondo systems, one remarkable phenomenon emerging from Kondo effect is the possible realization of a coherent macroscopic Fermi liquid ground state where localized quantum magnetic impurities contribute to the formation of non-local, i.e. Bloch waves, fermionic quasiparticle excitations. 
This fundamental issue is also very important for microscopic modelling of rare-earth compounds using 
renormalized quasi-particle band model~\cite{zwicknagl1992quasi} where $f-$electron levels may enter in the 
description of the Fermi surface. 

A standard approach for tuning the physical properties of concentrated Kondo systems is based on applying mechanical or chemical pressure and invoking Doniach's argument~\cite{doniach1977s}. Indeed, magnetic order is stabilized when the strength of the Kondo interaction, i.e. the corresponding $T_K$, is relatively small. As a consequence, very rich pressure-temperature phase diagrams are obtained in heavy-fermion Kondo $f-$electron materials, with a large diversity of unconventional quantum phases and behaviors~\cite{Coleman2015}. These include unconventional superconductivity~\cite{RevModPhys.81.1551,thalmeier2004unconventional} as observed, e.g. in CeCu$_2$Si$_2$~\cite{PhysRevLett.43.1892} or non-Fermi liquid behavior~\cite{lohneysen1994non}. In Kondo systems, the concentration of Kondo impurities $x$ and the electronic filling $n_c$ are other relevant parameters besides pressure.


Continuous tuning of $x$ can be realized experimentally in Kondo alloys heavy-fermions by isostructural substitution of magnetic rare-earth atom with a non-magnetic one. This induces remarkable changes in macroscopic physical properties. For example, Ce-La substitution in the series Ce$_x$La$_{1-x}$Cu$_6$~\cite{sumiyama1986coherent, onuki1987heavy} revealed an evolution from coherent dense Kondo lattice regime to dilute Kondo regime. In the series Ce$_x$La$_{1-x}$Ni$_2$Ge$_2$~\cite{PhysRevLett.108.066405} and Ce$_x$La$_{1-x}$PtIn~\cite{ragel2009dilution} additional non-Fermi liquid (nFL) behavior is observed along with dense-dilute transition whereas in the series Ce$_x$La$_{1-x}$FePO~\cite{chen2018spin} coexistence of spin glass with dilute Kondo regime is observed. Substitution of Ce (Kondo) by La (non-Kondo) can also produce antiferromagnetic (AF) to paramagnetic (PM) transitions in Ce$_x$La$_{1-x}$Cu$_2$Ge$_2$~\cite{PhysRevLett.114.236601} as well as PM to AF phase in Ce$_x$La$_{1-x}$Ru$_2$Si$_2$~\cite{aoki2014fermi}. Other types of rare-earth substitutions can also result in rich physics like Fermi-surface instabilities in unconventional superconductors like Ce$_x$Yb$_{1-x}$CoIn$_5$~\cite{PhysRevB.85.245119} 
and Nd$_{2-x}$Ce$_{x}$CuO$_4$~\cite{PhysRevLett.105.247002}. 

Multiple experimental and theoretical techniques were also used in order to explore the effect of $n_c$ for Kondo lattices (i.e., for $x=1$). Some of these studies revealed Doniach-like phase diagrams~\cite{lacroix1979phase, fazekas1991magnetic,Assaad1999,Peters2007}. 
An important attention was also focused on the issue of coherence in the paramagnetic phase of dense Kondo systems. 
This was first motivated by the question of `exhaustion' raised by Nozi\`eres~\cite{nozieres1985impuretes} 
regarding how a coherent Fermi liquid Kondo ground state can be formed when the concentration of Kondo impurities is larger than the concentration of conduction electrons. The robustness of a coherent Kondo state has then been investigated in the framework of the Kondo lattice model by several complementary theoretical approaches~\cite{lacroix1985some,lacroix1986coherence, Nozieres1998, 
PhysRevLett.85.1048,coqblin2000effect,PhysRevB.67.064417,PhysRevB.61.12799, PhysRevB.51.7429, PhysRevB.55.R3332, PhysRevLett.80.5168, PhysRevB.60.10782}. 
Meanwhile, photoelectron spectroscopy experiments~\cite{PhysRevLett.68.236, malterre1996recent, joyce1999photoemission} in Cerium and in Ytterbium heavy-fermion compounds found to be inconsistent with the predictions of the single impurity Kondo model. 

These works motivated further theoretical  investigations of the concomitant roles of $x$ and $n_c$ in the issue of coherence in Kondo alloys. A transition or a crossover between dilute and coherent dense Kondo Fermi liquid regimes had been suggested by various complementary methods including mean-field approximation and strong coupling expansions~\cite{burdin2007random, zu1988slave, burdin2019breakdown, burdin2013lifshitz, PhysRevB.75.132407,titvinidze2015strong,poudel2020phase}, Monte Carlo simulations~\cite{watanabe2010ground, PhysRevB.81.113111, otsuki2010effect}, numerical renormalization group~\cite{PhysRevB.77.115125}, or local-moment approach~\cite{PhysRevB.88.195120,PhysRevB.90.235133}. A crucial difference between these two Kondo Fermi-liquid regimes can be obtained from strong coupling description: for $x<n_c$ each Kondo impurity captures one conduction electron, forming Kondo singlets, and the Fermi-liquid quasiparticles correspond to the remaining $n_c-x$ electrons. For $x>n_c$, all conduction electrons form Kondo singlets and the Fermi-liquid quasiparticles emerge from the degrees of freedom of the remaining $x-n_c$ unscreened Kondo impurities. Invoking a particle-hole transformation, doping with Kondo impurities may thus correspond to particle doping in the dense coherent regime characterized by a large Fermi-surface enclosing $x+n_c$ particles, while it may correspond to hole doping in the dilute regime, with a reduced Fermi surface enclosing $n_c-x$ particles~\cite{burdin2013lifshitz}.  

While these theoretical works predict a fundamental difference between dense and dilute Kondo alloys, experimental signatures of such possible transitions are still not clear or mostly indirect. In this article, we explore one important property that may distinguish dense from dilute systems: the contribution (dense coherent Kondo) or the non-contribution (dilute Kondo) of f-electrons to the formation of the Fermi surface (FS). Here, we investigate from a theoretical point of view the robustness and breakdown of coherence in Kondo alloys with a focus on the possible signatures in Photo-emission spectroscopy (PES) and in Angle Resolved Photo-emission spectroscopy (ARPES). These experimental technics have already proven to be very useful skills for investigating the physics of $f-$electron systems, including Kondo alloys~\cite{fujimori2016band, RevModPhys.92.011002}. 
For example, PES experiments revealed limitations of the single impurity models for describing dense Kondo systems~\cite{PhysRevLett.68.236, malterre1996recent, joyce1999photoemission}. 
ARPES experiments carried on varieties of Kondo lattice systems like CeRu$_2$Si$_2$~\cite{denlinger2001comparative, PhysRevLett.102.216401, PhysRevB.77.035118}, CeRu$_2$Ge$_2$~\cite{PhysRevLett.98.036405}, CeBi~\cite{PhysRevB.100.155110},CeNiSn~\cite{PhysRevB.100.045133} and YbRh$_2$Si$_2$~\cite{PhysRevLett.107.267601, PhysRevX.5.011028, guttler2019divalent} showed large Fermi surfaces due to the coherent participation of 4$f$ electrons. ARPES experiments also permitted a direct observation of dispersive Kondo resonance peaks in CeCoGe$_{1.2}$Si$_{0.8}$~\cite{PhysRevLett.100.176402} and surface and bulk hybridizations in antiferromagnetic Kondo lattice CeRh$_2$Si$_2$~\cite{patil2016arpes}. 

In our work, we extend the matrix  DMFT/CPA approach developed in ~\cite{burdin2007random} to compute and analyze PES and ARPES signals for Kondo alloys on a square lattice structure. We consider both paramagnetic Kondo phases and magnetic ordered ground states. We start this paper by detailing model, methods and approximations in section~\ref{section:model}, results in section~\ref{section:results} followed by conclusion in section~\ref{section:conclusion}.

\section{Model and method} \label{section:model}
We consider the Kondo alloy model (KAM) Hamiltonian: 
\begin{eqnarray}
H=\sum_{ij\sigma}\left({t}_{ij}-\mu\delta_{ij}\right)c_{i\sigma}^{\dagger}c_{j\sigma}
+J_{K}\sum_{i\in {\cal K}}{\bf S}_{i}\cdot{\bf s}_i~, 
\label{EqKondoAlloyHamiltonian}
\end{eqnarray}
where $c_{i\sigma}^{(\dagger)}$ denotes creation (annihilation) operator of electron with spin component $\sigma=\uparrow,\downarrow$ on site $i$ of a periodic lattice. In this work, we consider a bipartite lattice (square or Bethe) that we formally split into two sublattices A and B. 
We consider non-zero electronic intersite tunneling $t_{ij}\equiv t$ for nearest neighbors $i$ and $j$ only. $\delta_{ij}$ is the Kronecker symbol, and the chemical potential $\mu$ is determined by fixing the average electronic filling per site to $n_c$. 
The term on the right hand side represents local Kondo impurities on a fixed subset ${\cal K}$ of lattice sites, which have been randomly distributed with a site concentration $x$. The antiferromagnetic interaction $J_K$ couples local Kondo quantum spin $1/2$ operators ${\bf S}_i$ with the local density of spin of conduction electrons ${\bf s}_i$. 
In the following, ${\cal N}$ (non-Kondo) will refer to the subset of non-Kondo sites. 
In this model, each local Kondo spin describes a local 4$f^1$ electronic state (Ce based materials) or a 4$f^{13}$ hole 
state (Yb based materials) with fixed valence. 

\subsection{DMFT/CPA method \label{section:DMFTmethod}}
We take into account the randomness of the $\mathcal{K}$ sites distribution in the KAM using a DMFT/CPA approach~\cite{georges1996dynamical,burdin2007random}. 
First, the action related with the KAM Hamiltonian Eq.~(\ref{EqKondoAlloyHamiltonian}) is expressed as 
\begin{equation} \label{action1}
S  = - \sum_{ij \sigma} \int_0^\beta  d\tau c_{i \sigma}^\dagger(\tau) \left( {(\partial_\tau - \mu )\delta_{ij}
+  t_{ij}}\right) c_{j \sigma}(\tau) - \int_0^\beta  d\tau  J_K \sum_{i \in \mathcal{K}} \textbf{S}_i(\tau) \textbf{s}_i(\tau)
\end{equation}
where $\beta=1/T$ is the inverse temperature, and $\tau$ denotes imaginary time. 
Hereafter, we will be following the matrix-DMFT formalism as developed in ~\cite{burdin2007random} for Kondo and non-Kondo paramagnetic solutions and generalized in ~\cite{poudel2020phase} to take account the possibility of some magnetically commensurate ordered phases. 
Introducing the sub-lattice index $\alpha=$A, B and the site-nature index $a=\mathcal{K}$ or $\mathcal{N}$, 
the lattice problem is thus mapped onto four effective single site problems embedded with dynamical local baths 
$\Delta_{a\alpha}^{\sigma}(\tau )= 
\sum_{ij}t_{\alpha i} t_{j \alpha} \langle c_{i \sigma}(\tau) c_{j \sigma}^\dagger(\tau^\prime)\rangle^{(\alpha,a)}$ that can be expressed in terms of the cavity ($\alpha, a$) electronic Green's functions, i.e.,  
using the KAM action~(\ref{action1}) without the contribution of the local site ($\alpha, a$). 
Furthermore, we take into acount randomness by assuming that the local $\mathcal{K}$ or $\mathcal{N}$ configuration of each site is not correlated with the configuration of neighboring sites. Hence the dynamical local baths 
for $\mathcal{K}$ or $\mathcal{N}$ sites are identical, so $\Delta_{\mathcal{K}\alpha}^{\sigma}(\tau) = \Delta_{\mathcal{N}\alpha}^{\sigma} (\tau )\equiv \Delta_{\alpha}^{\sigma}(\tau )$. 
The local effective actions are expressed as 
\begin{align}
S_\alpha^\mathcal{N} &= -\sum_{\sigma}\int_0^\beta d\tau \int_0^\beta d\tau^\prime c_{\mathcal{N}\alpha \sigma}^\dagger(\tau) 
\mathcal{G}_{\alpha \sigma}^{-1} (\tau - \tau^\prime) c_{\mathcal{N} \alpha \sigma}(\tau^\prime)~, \label{actionloc:nonkondo}\\
S_\alpha^\mathcal{K} &= -\sum_{\sigma}\int_0^\beta d\tau \int_0^\beta d\tau^\prime c_{\mathcal{K}\alpha \sigma}^\dagger(\tau) 
\mathcal{G}_{\alpha \sigma}^{-1} (\tau - \tau^\prime) c_{\mathcal{K} \alpha \sigma}(\tau^\prime) - J_K \int_0^\beta d\tau  \textbf{S}_\alpha(\tau) \textbf{s}_\alpha(\tau)~,  \label{actionloc:kondo} 
\end{align}
where the Matsubara frequency $i\omega$ dependencies of the kernel and the bath are related as 
$\mathcal{G}_{\alpha \sigma}^{-1} (i \omega) \equiv i\omega + \mu -\Delta_{\alpha}^{\sigma} (i\omega)$. 
The local electronic Green's functions $G_{a\alpha}^\sigma(\tau)\equiv -\langle c_{a\alpha\sigma}(\tau)c_{a\alpha\sigma}^{\dagger}(0)\rangle$ can be computed from the dynamical bath. 
The relation for $\mathcal{N}$-site is explicit since the 
corresponding local action~(\ref{actionloc:nonkondo}) is quadratic: 
\begin{equation}
G_{\mathcal{N}\alpha}^{\sigma}(i\omega)  = \frac{1}{i\omega + \mu - \Delta_{\alpha}^{\sigma}(i\omega) }~.  \label{MO:GreenNK}
\end{equation}
Computing the local Green function for $\mathcal{K}$-site requires an appropriate impurity solver for the 
Kondo interaction term in action~(\ref{actionloc:kondo}). The precise impurity solvers that we employed for the results presented in this paper are described in \ref{appendix:MF_approximations}. 

Hereafter we present the specific self-consistent relations that we developed in~\cite{poudel2020phase} using the DMFT/CPA approach, allowing to compute the dynamical bath from the local electronic Green's 
functions. We consider two complementary examples of bipartite lattices: a Bethe lattice, and a square lattice. 
The Bethe lattice is very convenient from the computational point of view. Indeed, considering the limit of a large lattice coordination $Z$ and rescaling the electronic exchange $t\mapsto t/\sqrt{Z}$, the self-consistency DMFT relation for a Bethe lattice is explicit: 
$\Delta_{\alpha}^{\sigma}(i\omega)=xt^2G_{\mathcal{K}\bar{\alpha}}^{\sigma}(i\omega) 
+(1-x)t^2G_{\mathcal{N}\bar{\alpha}}^{\sigma}(i\omega)$, where $\bar{\alpha}$ denotes the complementary sublattice of $\alpha$. However, while considering a Bethe lattice structure is very useful for computing local dynamical correlations, some specificities of the underlying periodic lattice and its reciprocal lattice are missing, that are crucial for describing ARPES. In this work we thus considered also the DMFT approach 
in its most general framework, having in mind a square lattice as bipartite structure.  
We introduce the ($2\times 2$) local Green function matrix: 
\begin{equation}
\mathbf{G}_{loc,\alpha}^{\sigma}(i\omega)  =  \begin{pmatrix}
x G_{\mathcal{K}\alpha}^{\sigma} (i\omega) & 0 \\
0 & (1-x) G_{\mathcal{N}\alpha}^{\sigma}  (i\omega)\\
\end{pmatrix}~, 
\label{equation:Glocmatrix}
\end{equation}
and the disorder-averaged site-dependent Green function matrix which preserves the sublattice A/B
symmetries : 
\begin{equation}
\mathbf{G}_{ij}^{\sigma}(i\omega)  =  \begin{pmatrix}
G_{ij\sigma}^{\mathcal{K}\mathcal{K}} (i\omega) & G_{ij\sigma}^{\mathcal{K}\mathcal{N}} (i\omega) \\
~~&~~\\
G_{ij\sigma}^{\mathcal{N}\mathcal{K}} (i\omega) & G_{ij\sigma}^{\mathcal{N}\mathcal{N}} (i\omega) \\
\end{pmatrix}~.   
\label{equation:Gijmatrix}
\end{equation}
The matrix elements are the electronic correlations $G_{ij\sigma}^{ab}(\tau)\equiv -\langle c_{i\sigma}(\tau)c_{j\sigma}^{\dagger}(0)\rangle$ that are firstly (i.e., before disorder-average) computed formally with the disordered KAM Hamiltonian~(\ref{EqKondoAlloyHamiltonian}) assuming the site-natures $a$ and $b$ for $i$ and $j$ respectively. 
We also introduce the momentum representation: 
$\mathbf{G}^{\sigma}_{\bf kk'}(i\omega)\equiv
\frac{1}{N}\sum_{ij}e^{i{\bf k}'{\bf R}_j-i{\bf k}{\bf R}_i)}\mathbf{G}_{ij}^{\sigma}(i\omega)$, 
where $N$ is the number of lattice sites. 
In order to take into account the possibility of ordering that may break lattice symmetry with Néel order, 
we introduce the wavevector $\mathbf{Q}=(\pi,\pi)$. 
We then define the ($4\times 4$) matrix $\mathbf{\bar{\bar{G}}^\sigma_\mathbf{kk'}} 
\equiv \begin{pmatrix}
\mathbf{G}^\sigma_{\bf kk'} & 
\mathbf{G}^\sigma_{\bf kk'+Q}\\
\mathbf{G}^\sigma_{\bf k+Qk'} & 
\mathbf{G}^\sigma_{\bf k+Qk'+Q}
\end{pmatrix}$. 
Exploiting symmetric properties of the Green functions for the specific cases of paramagnetic Kondo, ferromagnetic and staggered antiferromagnetic orders, and invoking the DMFT diagrammatic expansion, we obtain the following relation: 
\begin{equation}
\label{appendix-equation:Greenenk}
\mathbf{\bar{\bar{G}}^\sigma_\mathbf{kk'}}(i\omega) = \delta_{\mathbf{k k'}}
\left( \begin{pmatrix}
	\mathbf{\Pi}_+^\sigma (i\omega)& \mathbf{\Pi}_-^\sigma (i\omega)  \\
	\mathbf{\Pi}_-^\sigma (i\omega) & \mathbf{\Pi}_+^\sigma (i\omega)
	\end{pmatrix}^{-1} - 
	\begin{pmatrix}
	\mathbf{E_k} & 0 \\
	0 & \mathbf{E_{k+Q}}
	\end{pmatrix} \right)^{-1},  
\end{equation}
where $\mathbf{\Pi}_\pm^\sigma (i\omega)$ is a local propagator matrix which emerges in DMFT diagrammatic expansion and 
\begin{equation}
	\mathbf{E_k} = \epsilon_{\mathbf{k}} \begin{pmatrix}
	1 & 1\\
	1 & 1
	\end{pmatrix}~, 
\end{equation}
with the non-interacting dispersion 
$\epsilon_{\mathbf{k}} = 2t[\cos{(k_x)}+\cos{(k_y)}]$. The DMFT self-consistency is closed by expressing  
$\mathbf{G}_{loc,\alpha}^{\sigma}(i\omega)=\mathbf{G}_{ii}^{\sigma}(i\omega)$ with $i$ on sublattice $\alpha$. 
When solving this identity in a DMFT self-consistent scheme, the local Green function is supposed to be known from 
Eq.~(\ref{equation:Glocmatrix}) and having solved the local effective action problems given by 
Eqs.~(\ref{actionloc:nonkondo}) and~(\ref{actionloc:kondo}). 
In turn, the local effective bath which is involved in these actions can be obtained with algebraic matrix inversion as follows: 
\begin{equation} 
\mathbf{\Delta}_{\alpha}^{\sigma}(i\omega) \equiv  [\mathbf{\Pi}_{\alpha}^{\sigma}(i\omega)]^{-1} - [\mathbf{G}_{loc,\alpha}^{\sigma}(i\omega)]^{-1}~,    \label{DMFT:equationBain1}
\end{equation}
where the matrices $\mathbf{\Pi}_{A}^{\sigma}(i\omega)=
\mathbf{\Pi}_{+}^{\sigma}(i\omega)+\mathbf{\Pi}_{-}^{\sigma}(i\omega)$ 
and $\mathbf{\Pi}_{B}^{\sigma}(i\omega)=
\mathbf{\Pi}_{+}^{\sigma}(i\omega)-\mathbf{\Pi}_{-}^{\sigma}(i\omega)$ 
can be obtained from the local Green function by invoking the following identity: 
\begin{equation}\label{DMFT:local_green_fonction_alliage}
\mathbf{G}_{loc,\alpha}^{\sigma}(i\omega) = \frac{1}{N_{RBZ}} \sum_{\mathbf{k} \in RBZ} 
([\mathbf{\Pi}_{\alpha}^{\sigma}(i\omega)]^{-1} -  \mathbf{E_k} \mathbf{\Pi}_{\bar{\alpha}}^{\sigma}(i\omega) \mathbf{E_k})^{-1}~.  
\end{equation}
Here, $N_{RBZ}$ is the volume of the reduced Brillouin zone that corresponds to one sublattice A/B. 
Finally, the dynamical local bath $\Delta_{\alpha}^{\sigma}(i\omega)$ 
is deduced from the diagonal matrix elements: 
\begin{equation}
\Delta_{\alpha}^{\sigma}
=\left( \mathbf{\Delta}_{\alpha}^{\sigma}\right)_{\mathcal{K}}
=\left( \mathbf{\Delta}_{\alpha}^{\sigma}\right)_{\mathcal{N}}~.  
\label{DMFT:equationBain2}
\end{equation}
Once self-consistent solution of the DMFT effective problem is obtained, physical quantities can be computed from the KAM 
and provide predictions for photo-emission experiments in Kondo alloy materials. 
For example, the average electronic local density of states (d.o.s.) per spin component can be computed as: 
\begin{eqnarray}
\rho(\omega)&=& -\frac{1}{4\pi}\sum_{\sigma,\alpha}{\cal I}m[ xG_{\mathcal{K}\alpha}^{\sigma}(\omega)
+(1-x)G_{\mathcal{N}\alpha}^{\sigma}(\omega)]~. \nonumber\\
&&\label{ARPESequation}
\end{eqnarray}
The d.o.s. can be measured by PES, and it may be related with the ARPES signal by a sum rule (see section~\ref{Section:inelasticARPES}).

\subsection{Angle Resolved Photoemission spectroscopy (ARPES) in a paramagnetic Kondo phase}
In this section we present the adaptation of the DMFT/CPA method for modelling ARPES experimental measurements 
on Kondo alloy materials. For the sake of clarity, we assume here that the system is in a paramagnetic phase and we skip sublattice and spin indices $\alpha$ and $\sigma$, focussing on the specificities of the random distribution of ${\cal K}$ and ${\cal N}$ sites on a periodic square lattice. 
The system is disordered, however the signal which is measured by ARPES is the spectral weight 
${\cal A} ({\bf k},\omega)=-\frac{1}{\pi}{\cal I}m[G_{Alloy}({\bf k},\omega+i0^{+})]$
which is associated with the intersite one-electron correlation function 
$G_{Alloy}({\bf k},i\omega)=\frac{1}{N}\sum_{ij}e^{i{\bf k}({\bf R}_j-{\bf R}_i)}{G}_{ij}(i\omega)$.  
Here $\mathbf{k}$ denotes the momentum in the Brillouin zone of the square lattice. The site-dependent electronic Green function 
${G}_{ij}(i\omega)$ is in principle dependent on the disorder configuration, i.e., on the specific distribution of 
${\cal K}$ and ${\cal N}$ sites. However, the ARPES signal is rather related with $G_{Alloy}({\bf k},i\omega)$ which 
involves a sum over all lattice sites $i$ and $j$. We assume that the disorder is self-averaged when performing 
this sum. In principle this is realized if the system is sufficiently big.  
Therefore, the ARPES spectral weight of a Kondo alloy system can be obtained from the disorder-averaged Green-function computed from the KAM using the DMFT/CPA method: 
\begin{equation}
G_{Alloy}({\bf k},i\omega) = G_{\bf k}^{\mathcal{K}\mathcal{K}}(i\omega)
+ G_{\bf k}^{\mathcal{K}\mathcal{N}}(i\omega) + G_{\bf k}^{\mathcal{N}\mathcal{K}}(i\omega)
+ G_{\bf k}^{\mathcal{N}\mathcal{N}}(i\omega)~,  \label{onebody:def}
\end{equation}
where $G_{\bf k}^{ab}(i\omega)$ denote the elements of $G_{\bf kk}(i\omega)$ which is 
the momentum representation of the Green function matrix (\ref{equation:Gijmatrix}). This expression reflects the fact that ARPES on a Kondo alloy involves all one-electron excitations irrespectively from the Kondo or non-Kondo nature of the sites where a conduction electron is created or annihilated. 
Invoking the DMFT self-consistent condition $\sum_{\bf k}G_{\bf k}^{\mathcal{N}\mathcal{K}}(i\omega)=
\sum_{\bf k}G_{\bf k}^{\mathcal{K}\mathcal{N}}(i\omega)=0$ which reflects the non-ambiguity of the local nature of a given site, we can check that the sum rule is satisfied between the ARPES signal and the local density of states: 
$\rho(\omega)=\frac{1}{N}\sum_{\bf k}{\cal A} ({\bf k},\omega)$. 

In a paramagnetic phase, the propagator involved in the DMFT/CPA self-consistency becomes independent from sublattice and spin indices: $\mathbf{\Pi}_{\alpha}^{\sigma}(i\omega)\equiv \mathbf{\Pi}(i\omega)$. 
The Green function matrix can then be cast into the same form as in (~\cite{burdin2007random}): 
\begin{eqnarray}
G_{\bf kk}(i\omega)= \left( \mathbf{\Pi}^{-1}(i\omega)-{\bf E}_{\bf k}\right)^{-1}~. 
\label{Greenkomegaparamagn}
\end{eqnarray}
With this effective description, the ARPES spectra are related with the dispersion relation 
$det\left( \mathbf{\Pi}^{-1}(\omega)-{\bf E}_{\bf k}\right)=0$, which may lead to the possible emergence of two branches (in momentum) since $\mathbf{\Pi}$ and ${\bf E}_{\bf k}$ are $2\times 2$ matrices. Such a two branches 
ARPES signature is, for example, obtained in the mean-field description of a periodic Kondo lattice ($x=1$) in terms of effective hybridization between conduction electrons and localized electronic $f-$levels. However the dispersion may also be single branch, like, for example, in the non-interacting case where $J_K=0$ as well as in the dilute limit $x\to 0$. 
The DMFT/CPA method applied here for the KAM provides a unique framework for describing both this single and double branch scenarii. 

In order to distinguish the transitions and crossovers between dense and dilute Kondo alloys, a useful quantity will be the effective mass that could be related with ARPES signal. 
First, we introduce the Kondo local self-energy as: 
\begin{eqnarray}
\Sigma_K(i\omega)\equiv \frac{1}{G_{\mathcal{N}}(i\omega)}-\frac{1}{G_{\mathcal{K}}(i\omega)}~. 
\label{equation:kondoselfenergy}
\end{eqnarray}
Invoking the Eqs.~(\ref{DMFT:equationBain1}-\ref{DMFT:equationBain2}) and with some straightforward algebra, the local propagator matrix  involved in the Eq.~\ref{Greenkomegaparamagn} can be expressed explicitly in terms of 
$\Delta (i\omega)$ and $\Sigma_K(i\omega)$ as: 
\begin{eqnarray}
	\mathbf{\Pi}^{-1}(i \omega) = \begin{pmatrix}
	\frac{i \omega + \mu - \Sigma_{K}(i \omega) - (1-x) \Delta(i \omega) }{x}  & \Delta(i \omega)  \\
	~~&~~\\
	\Delta(i \omega) & \frac{i \omega + \mu - x \Delta(i \omega) }{1-x}
	\end{pmatrix}~, \nonumber\\
	~~
\end{eqnarray} 
Using this expression together with Eqs.~(\ref{onebody:def}-\ref{Greenkomegaparamagn}), the one-body 
disorder-averaged Green-function involved in ARPES measurements can be expressed as: 
\begin{eqnarray}
G_{Alloy}({\bf k},i\omega)=\frac{1}{i\omega+\mu-\Sigma_{Alloy}(i\omega)- \epsilon_{\mathbf{k}}}~,  \label{EquationGkmoyenne}
\end{eqnarray}
where the effective self-energy is defined as: 
\begin{equation} \label{self-energy}
\Sigma_{Alloy}(i\omega) = \frac{x}{\frac{1}{\Sigma_{K}(i\omega)} - \frac{(1-x)}{i\omega+\mu-\Delta (i\omega)}}~. 
\end{equation}
We can easily verify that this expression reproduces standard results at several limits. At extreme dilute limit ($x\to 0$), 
the effective self-energy $\Sigma_{Alloy}$ vanishes linearly with $x$ and the non-interacting electronic Green function is recovered 
$G_{Alloy}({\bf k},i\omega)\to1/(i\omega+\mu- \epsilon_{\mathbf{k}})$, 
with its single branch signature. In the Kondo lattice limit $x\to1$, $\Sigma_{Alloy}\to\Sigma_{K}$ and a two-branches signature can be obtained within a Kondo mean-field approximation that introduces a pole singularity 
for $\Sigma_{K}(i\omega)$. 
We may estimate from expression~(\ref{self-energy}) that $\Sigma_{Alloy}\approx x \Sigma_{K}$ when 
$x\approx 1$. This linearity with Kondo spin concentration is approximately valid only at high and low concentrations. This suggests that the two-branches signatures of the Kondo lattice coherent state should be preserved upon slight Kondo dilution. Similarly, the single branch signature should also be maintained in the very diluted limit. However, this linearity is not valid anymore at intermediate concentrations and the breakdown of coherence that was predicted elsewhere around $x=n_c$~\cite{burdin2013lifshitz} might lead to ARPES signatures. In order to address this issue more quantitatively, we introduce the effective mass $m^*$ as:  
\begin{equation} \label{Equation:mass_effective}
	\frac{m^*}{m_0} = 1 - \left.{\frac{d {\cal R} e (\Sigma_{Alloy}(\omega))}{d \omega}}\right|_{\omega=0}~, 	
\end{equation}
where  $m_0$ is the mass of electron without interaction. 

\subsection{Impurity solver used for DMFT/CPA: mean-field approximations} \label{DMFT:Impurity_solver}
One key step in solving the DMFT/CPA self-consistent equations as described in section~\ref{section:DMFTmethod} is the 
resolution of the many-body problem defined by the local effective action on a Kondo site, Eq.~(\ref{actionloc:kondo}). 
Here, we employ mean-field approximations as local impurity solvers for describing non-Kondo purely magnetically ordered phases and pure Kondo paramagnetic phases. Indeed, the aim is to investigate the effect of Kondo alloying with two complementary perspectives: {\it i)} Describing the Fermi-surface folding of antiferromagnetic N\'eel ordered phases, {\it ii)} Describing the possible connection between the one-branch versus two-branches ARPES signatures of the Kondo alloys. 
We thus use the Weiss mean-field approximation for addressing point {\it i} (see \ref{appendix:MF_approximation_Weiss}), and we use the Kondo mean-field approximation for addressing point {\it ii} (see \ref{appendix:MF_approximation_Kondo}) since it is able to describe the one-branch and two-branches ARPES in the dilute limit ($x=0$) and for the coherent Kondo lattice ($x=1$) respectively. The competition between these various states as predicted by Doniach~\cite{doniach1977s} and the resulting ground state phase diagram of the KAM was analyzed in details in~\cite{poudel2020phase} as function of $x$, $n_c$, and $J_K$. Here, we want to focus onto the ARPES and PES signatures of these different phases that may emerge from the KAM~\cite{poudel2020phase}. \\
In our mean-field approach, the Kondo crossover at Kondo temperature $T_K$ is replaced by a transition where $f$ and $c$-electrons locally hybridize. The corresponding mean field equation for $T_K$ is
\begin{equation}
	\frac{2}{J_K} = \int_{-\infty}^{+\infty} \frac{\rho_0(\omega+\mu)}{\omega}\tanh(\frac{\omega}{2T_K})d\omega
\end{equation}
where $\rho_{0}(\omega)$ is non-interacting density of states. It is worthwhile noting here that $T_K$ is independent on $x$, as discussed in\cite{burdin2007random}.  In the following, the strength of the Kondo interaction will be adjusted by comparing $T_K$ with the non-interacting electronic bandwidth $W$. 



\section{Results}~\label{section:results}
In this section, we present and discuss our numerical results focusing on photo-emission signatures obtained for the KAM on a square lattice. In order to confirm the validity of our results, we also checked if some effects might be related to the specific square lattice-structure. For that, we also performed DMFT self-consistency over Bethe lattice (BL). No qualitative significant differences were found between Bethe and square lattices regarding local self-energies in the paramagnetic phases. \\
For both lattices, self-consistent solutions for the dynamical bath from the DMFT equations (\ref{DMFT:equationBain1},~\ref{DMFT:local_green_fonction_alliage}) and the local mean-field approximations (~\ref{ferro:MFequation1},~\ref{ferro:MFequation2},~\ref{ferro:MFequation3} or~\ref{Kondo:MFequation1},~\ref{Kondo:MFequation2},~\ref{Kondo:MFequation3}) were obtained numerically using standard optimization methods that we adapted for this specific problem. In this article, we considered pure paramagnetic Kondo, ferromagnetic (F) and staggered antiferromagnetic (AF) phases and the results below are presented for 2D square lattice with electronic bandwidth $W\equiv 8t$. 

\subsection{ARPES and electronic density of states \label{Section:inelasticARPES}}

\begin{figure}
	\centering	
\includegraphics[width=0.20\textwidth]{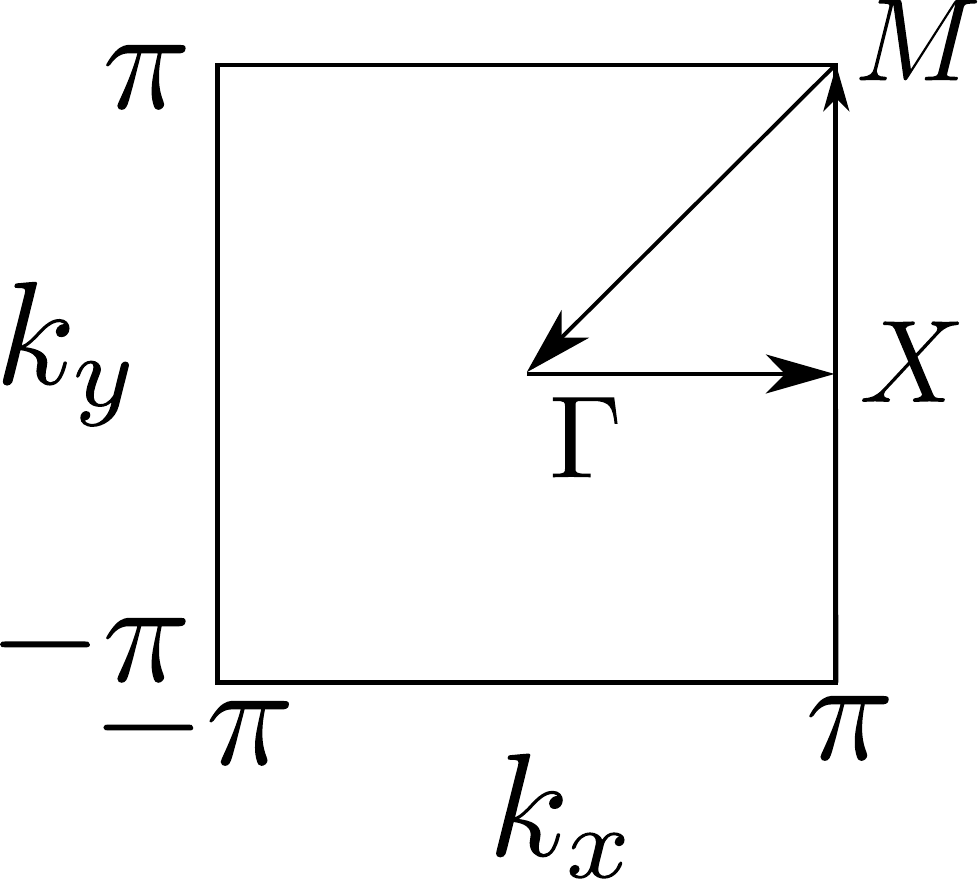} 
	\caption{First Brilloin zone of the square lattice, with indications of the points $\Gamma=(0,0)$, $X=(0,\pi)$, and $M=(\pi,\pi)$. 
}
	\label{Figure:2DBZ}
\end{figure} 

In this section, we analyze the ARPES signals obtained through the spectral function ${\cal A} ({\bf k},\omega)$ and the local electronic density of states $\rho(\omega)=\frac{1}{N}\sum_{\bf k}{\cal A} ({\bf k},\omega)$ with a focus on the specific signatures of the transitions and crossovers in the paramagnetic Kondo phases. In particular we address the issues of one-branch versus multi-branches dispersions in ARPES along with gap-less, pseudogapped, or gapped local electronic density of states. 

\begin{figure}
	\centering
	\vspace{-10pt} 
	\includegraphics[width=\textwidth]{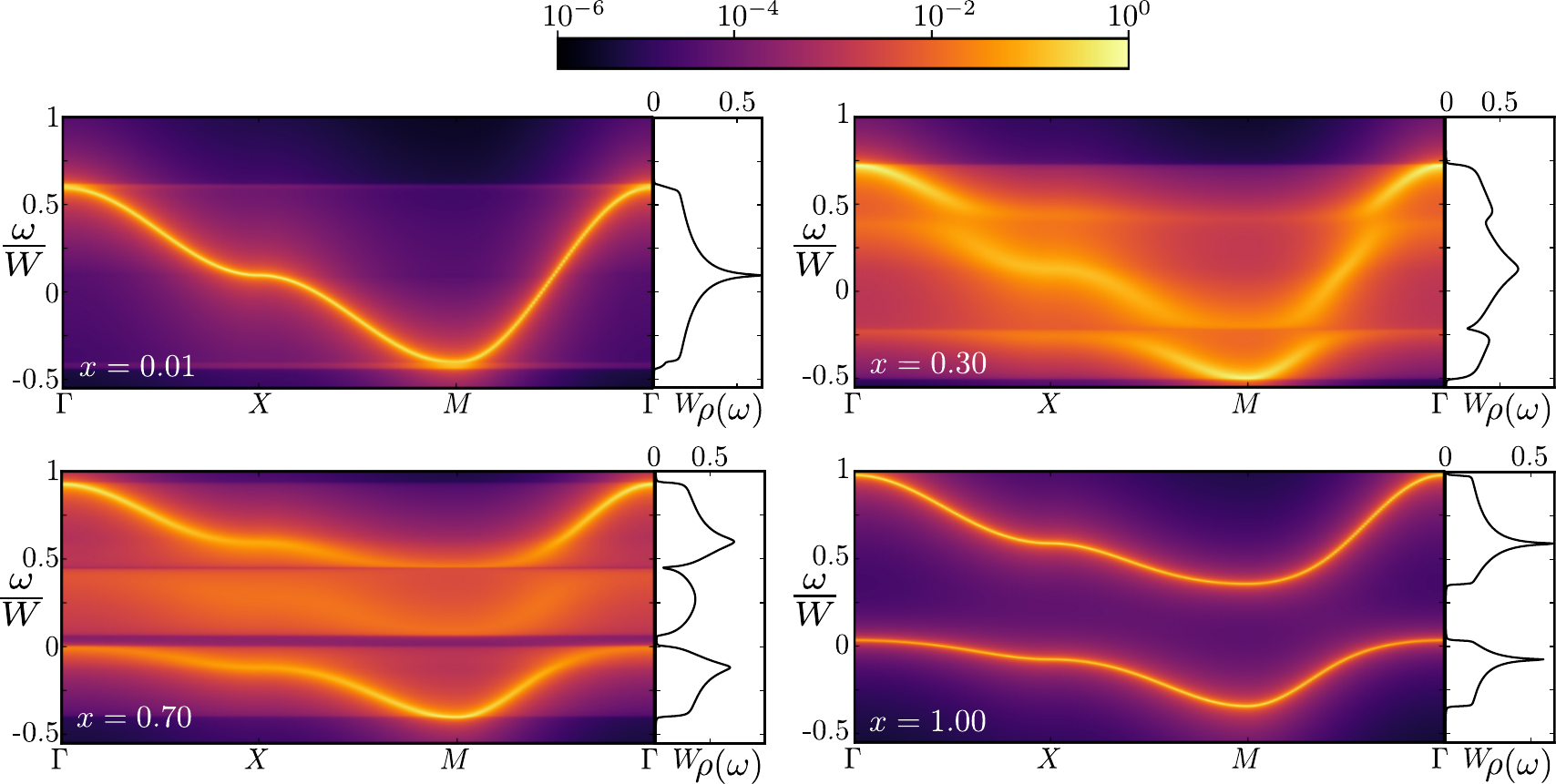}
	\caption{ \small{ARPES spectra for $n_c$ = 0.70 at relatively strong coupling $T_K/W$ = 0.169 for Kondo impurity concentrations $x=0.01$, $0.30$, $0.70$, and $1.00$. The wavevector ${\bf k}$ axis corresponds to the high symmetry lines $\Gamma-X-M-\Gamma$ in the square lattice first Brillouin zone (see figure~\ref{Figure:2DBZ}). The corresponding electronic density of states $\rho(\omega)$ is plotted on the right side.
	}\label{Figure:inelasticARPESFort}}
\end{figure}

\begin{figure}
	\centering
	\vspace{-10pt} 
	\includegraphics[width=\textwidth]{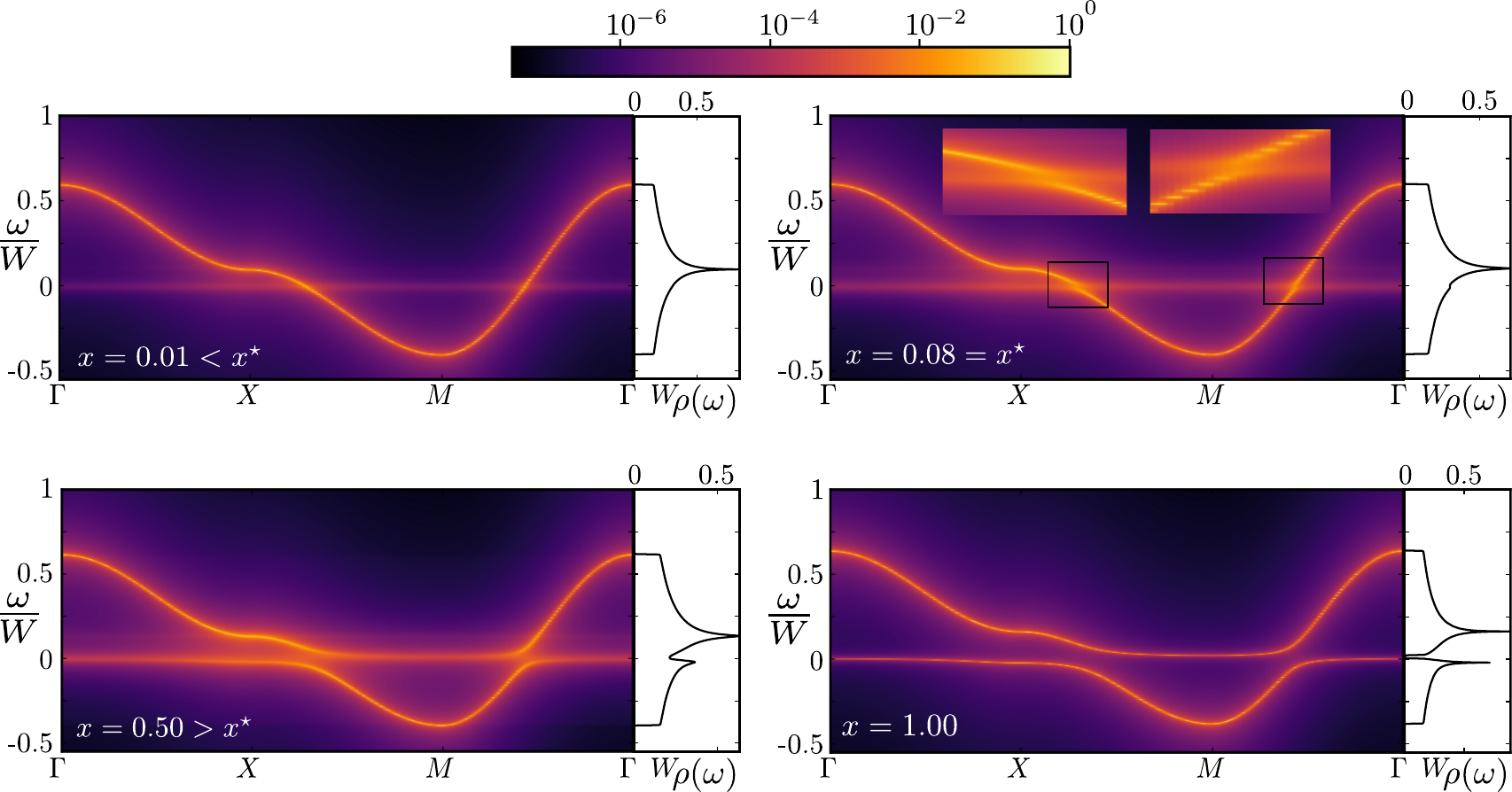}
	\caption{ \small{ARPES spectra for $n_c$ = 0.70 at relatively small coupling $T_K/W$ = 0.019 for Kondo impurity concentrations $x=0.01$, $0.08$, $0.50$, and $1.00$ with $x^\star= 0.08$.
	The wavevector ${\bf k}$ axis corresponds to the high symmetry lines $\Gamma-X-M-\Gamma$ in the square lattice first Brillouin zone (see figure~\ref{Figure:2DBZ}). The corresponding electronic density of states $\rho(\omega)$ is plotted on the right side.  }\label{Figure:inelasticARPESFaible}
	}
\end{figure}

We solved the self-consistent equations for various electronic fillings and we obtained similar qualitative results in all cases. Therefore here, we choose $n_c = 0.70$ to present an overview of different situations depending on $x$ and $T_K$. Overall we identify two different scenarii depending on the strength of the Kondo interaction: the relatively high $T_K$ regime (see figure~\ref{Figure:inelasticARPESFort}) and the relatively small $T_K$ regime (see figure~\ref{Figure:inelasticARPESFaible}), typically separated by a threshold corresponding to $T_K$ around $W/10$.
In both cases, the dilute Kondo limit ($x\ll 1$) and the dense Kondo limit ($x=1$) generate usual results. The dilute Kondo limit ($x\ll 1$) reproduces the non-interacting electronic structure, which is characterized here by a well defined single branch electronic dispersion and a density of states with a van-Hove singularity usual for the square lattice. The Kondo lattice limit ($x=1$) also presents universal signatures with two branches resulting from the effective hybridization between conduction electrons and the local levels associated with Kondo spins. The resulting density of states also presents a gap. For intermediate concentrations $x$, the situation depends on the strength of the Kondo interaction.
\subsubsection{Evidence for a Lifshitz-like transition in the Kondo phase at large $T_K$ \\} 
The numerical results obtained for various Kondo impurity concentrations at relatively large Kondo coupling~$T_K/W=0.169$ are presented in the figure~\ref{Figure:inelasticARPESFort}. In the intermediate concentration regime, as long as $x>n_c$, the dilution of Kondo impurities does not close the hybridization gap characterizing the coherent dense regime. Since we have fixed $n_c < 1$ in the dense case, the Fermi level is inside the lower band. When decreasing $x$, a third band starts to be formed inside the gap and a transition from dense to dilute Kondo regimes occurs at $x=n_c$. This transition is marked by the shift of the Fermi level from the lower band to the third band. Upon further dilution of Kondo impurities in the regime $x < n_c$, the hybridization gap is filled giving rise to three branches structure separated by two pseudogaps.

Let us consider at first $x>n_c$ regime in order to understand the formation of the third band and the breakdown of the coherence of the Kondo impurities. In this dense regime and at strong Kondo coupling, the KAM can be mapped onto an effective Hubbard model where quasiparticles are the unscreened Kondo impurities~\cite{PhysRevB.30.5383,lacroix1985some, bastide1987d, PhysRevB.46.13838}. The corresponding ``Coulomb repulsion'' is in this case of the order $T_K$ which is very large compared to $W$. Apart from half-filling which corresponds to a Kondo insulator for $x=1$, the system is a strongly correlated metal. Considering particle-hole general symmetry, let us depict a situation where the effective Fermi level is inside the lower band. The gapped local d.o.s characterizing Kondo lattices at large $T_K$ reflects the two Hubbard bands separated by an energy of the order of $T_K$ and the states in the upper Hubbard band correspond to singlet-triplet excitations. Dilution of Kondo atoms in the dense regime ($n_c<x<1$) changes the number of carriers and the Fermi level gets closer to the upper edge of the effective Hubbard band. The transition realized at $x=n_c$ for strong Kondo interaction may thus be analogous to a doping-induced Mott transition \cite{RevModPhys.70.1039}, which also presents the formation of a quasiparticule peak inside the Hubbard gap. However, the effective model is different for $x<n_c$: in this dilute regime and for strong Kondo interaction, quasiparticles emerge from the supernumerary conduction electrons which do not form Kondo singlets. The third central band may be associated with the motion of these conduction electrons on the non-Kondo sites. Further, the states in the fully occupied lower band represent the electrons forming singlets on Kondo sites whereas the upper unoccupied band corresponds to excitations of a second electron on a Kondo site.


\subsubsection{Evidence for a new critical concentration at low $T_K$ \\}
We now focus on the ARPES signal obtained for the relatively small Kondo coupling case. The figure~\ref{Figure:inelasticARPESFaible} illustrates the results obtained for $T_K/W=0.019$. We observe that the two branches structure characterizing the dense Kondo state is preserved upon dilution even for $x<n_c$. Furthermore, the effect of disorder-related decoherence is maximum at around $x=n_c$ which leads to a broadening of the branches as well as a reduction of the quasiparticle lifetime. This maximum decoherence also results in a partial filling of the hybridization gap leaving a pseudogap near the Fermi level. Upon further dilution of Kondo atoms, the two ARPES branches merge to form a single branch structure along with the disappearance of the pseudogap. This occurs at a critical concentration $x^\star$ which depends on the strength of the Kondo interaction. We find that $x^\star \ll n_c$ at very small $T_K$, and $x^\star \to n_c$ when $T_K$ approaches around $W/10$.



\subsection{Fermi surfaces} \label{section:elasticARPES}
\begin{figure}
    \centering	
    \includegraphics[width=0.40\textwidth]{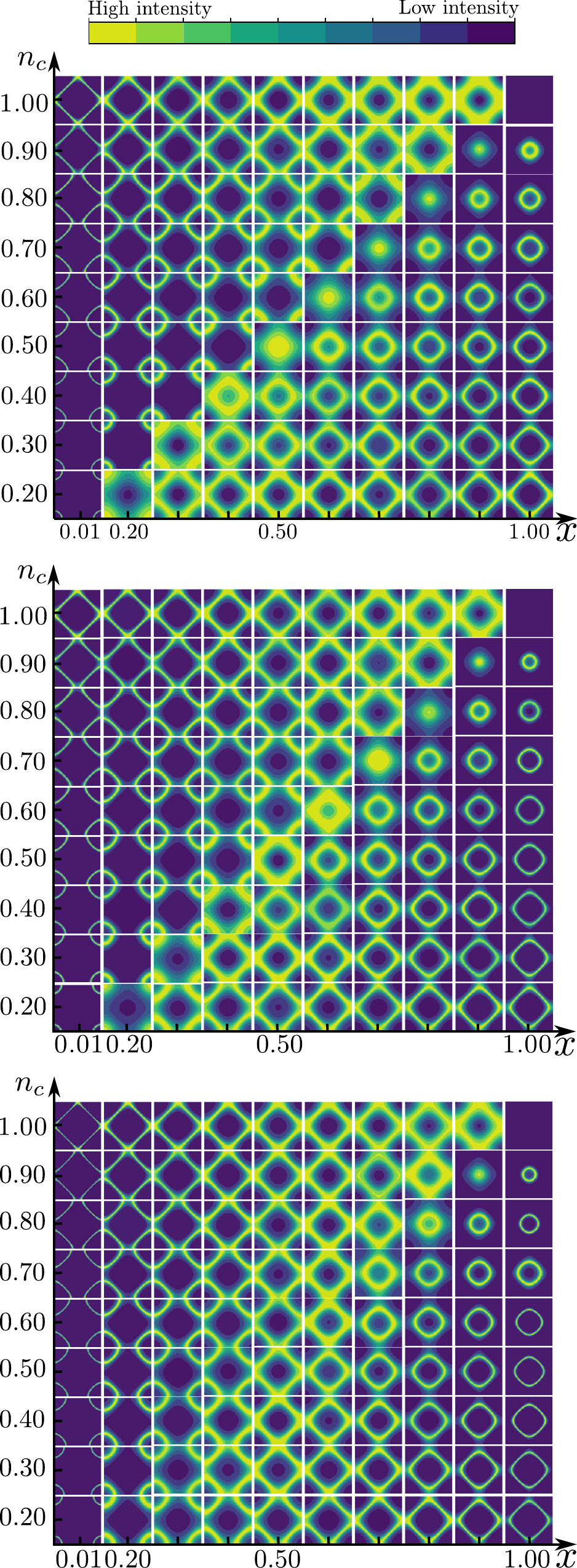}
	\caption{FS spectra assuming a Kondo paramagnetic ground state for $x=0.01$,  $0.20$, $0.30$, $0.40$, $0.50$, $0.60$, $0.70$, $0.80$, $0.90$, $1.00$ (from left to right) and $n_c=0.20$, $0.30$, $0.40$, $0.50$, $0.60$, $0.70$, $0.80$, $0.90$, $1.00$ (from bottom to top). 
Each square corresponds to the First Brillouin zone of the square lattice (see figure~\ref{Figure:2DBZ}). 
From top to bottom, $T_K/W=0.175$, $0.082$, and $0.0058$: the Lifshitz-like transition around $x=n_c$ is observed for a sufficiently strong Kondo interaction, and it becomes a gradual crossover for smaller values of the interaction. 
 }
	\label{Figure:ARPESElasticKondoPhase}
\end{figure}

In this section we analyze the experimental signatures in the Fermi surfaces obtained from the ARPES signal ${\cal A} ({\bf k},i\omega =i0^+)$. Figure~\ref{Figure:ARPESElasticKondoPhase} depicts the Fermi surface spectra computed for a paramagnetic Kondo ground state within broad ranges of $x$ and $n_c$. For all values considered for the Kondo interaction and the electronic filling, we find that the Fermi surface of the Kondo lattice ($x=1$) is large and it includes the contributions from both the conduction electrons and the Kondo spins. This universal feature is in good agreement with previous theoretical and experimental results~\cite{fujimori2016band, RevModPhys.92.011002, denlinger2001comparative, PhysRevLett.102.216401, PhysRevB.77.035118, PhysRevLett.98.036405, PhysRevB.100.155110, PhysRevB.100.045133, PhysRevLett.107.267601, PhysRevX.5.011028, guttler2019divalent}. It can be well understood in terms of Luttinger theorem, which  stipulates that all fermionic degrees of freedom participate in the formation of the Fermi liquid ground state. 

\subsubsection{Fermi surface in the Kondo phase: coherence breakdown \\} We now focus on the possible breakdown of this coherent Kondo lattice state when decreasing $x$. For a relatively strong Kondo interaction ($T_K/W=0.175$) we observe the Lifshitz-like transition at $x=n_c$ that was predicted in~\cite{burdin2013lifshitz}. In this case, the ARPES signal corresponds to relatively long lifetime quasiparticles with a well defined Fermi surface (excepted at the transition). 
The volume of the Fermi-surface shrinks when increasing $x$ in the dilute regime $x<n_c$, and it increases with $x$ in the dense regime $x>n_c$. This feature is consistent with the fact that Kondo impurities behave as hole-dopant for $x<n_c$ and particle-dopant for $x>n_c$. 
For $T_K/W=0.082$, and $0.0058$, the evolution of the Fermi surface ARPES spectra with concentration $x$ is more gradual, and the Lifshitz-like transition around $x=n_c$ seems to become a crossover at lower values of $T_K$. Furthermore, the broadening of the Fermi surface spectra around the Fermi wavevectors is maximal around $x=n_c$ due to disorder-related decoherence.    

Fermi surface spectra provide apparent signatures of the Kondo lattice coherence breakdown transition at $x=n_c$, especially for systems with relatively strong Kondo interaction, corresponding to $T_K$ typically larger than $W/10$. This Lifshitz like transition separating dense and dilute Kondo phases at $x=n_c$ becomes a crossover for smaller $T_K$. At small $T_K$, we identified another critical concentration from the ARPES signals (see section~\ref{Section:inelasticARPES}) at $x^\star$ characterizing the merging of two ARPES branches. However, the Fermi surfaces at figure~\ref{Figure:ARPESElasticKondoPhase} do not present clear signatures of any characterized feature at or around this critical concentration $x^\star$. This is not surprising since the electronic excitation spectrum is not accessible from the Fermi surfaces.

\begin{figure}
	\centering	
\includegraphics[width=0.40\textwidth]{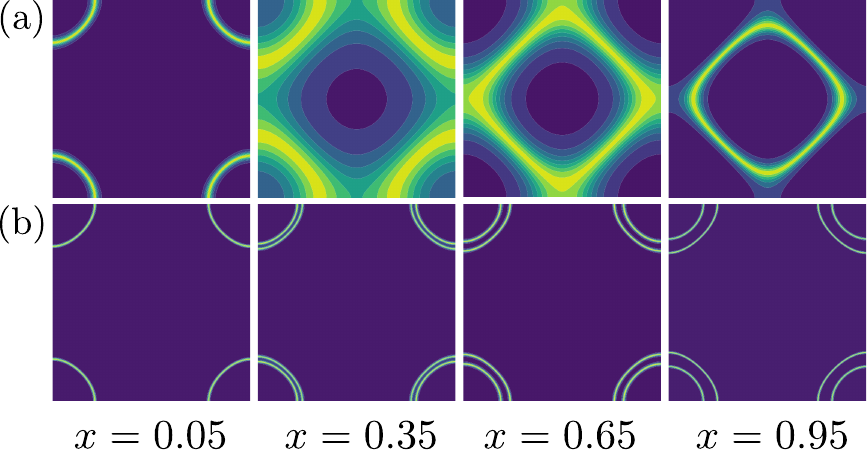} 
	\caption{Fermi surface spectra for $n_c=0.30$ and $x=0.05$, $0.35$, $0.65$ and $0.95$. 
	(a) $T_K/W=0.012$ with Kondo ground states: we observe signatures of a breakdown of coherence associated with a change of topology in the the Fermi-surface. (b) $T_K/W=0.0031$ with ferromagnetic ground states: we observe Zeeman splitting effect only.  
 }
	\label{Figure:ARPESElasticFerroKondo}
\end{figure}

\begin{figure}
	\centering	
\includegraphics[width=0.40\textwidth]{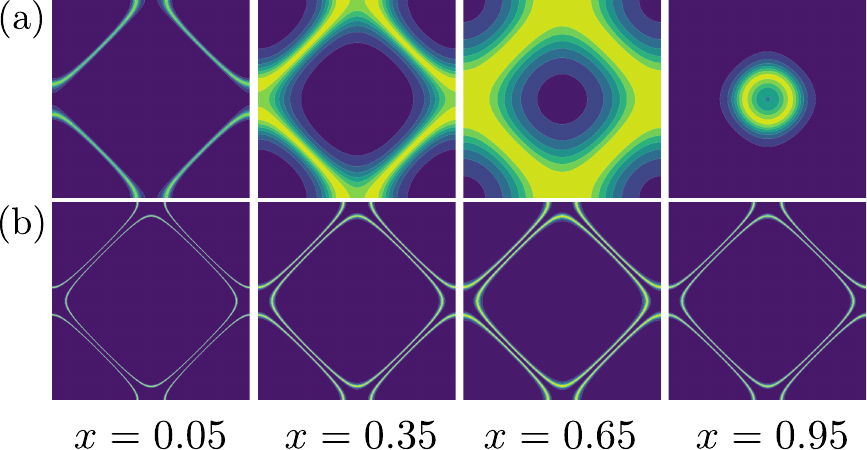} 
	\caption{Fermi surface  spectra for $n_c=0.90$, and $x=0.05$, $0.35$, $0.65$ and $0.95$.  
	(a) $T_K/W=0.0143$ with Kondo ground states: we observe signatures of a breakdown of coherence associated with a change of topology in the the Fermi-surface. (b) $T_K/W=0.0045$ with antiferromagnetic ground states: we observe 
	only the folding of the Fermi-surface which results from the staggered N\'eel ordering. 
 }
	\label{Figure:ARPESElasticAntiFerroKondo}
\end{figure}
\subsubsection{Fermi surface in the magnetically ordered phases\\} 
We also studied the Fermi surfaces in the magnetically ordered phases of Kondo alloys, as depicted in figure~\ref{Figure:ARPESElasticFerroKondo} (for $n_c=0.30$, where the small Kondo coupling ground state is ferromagnetic) and in figure~\ref{Figure:ARPESElasticAntiFerroKondo} (for $n_c=0.90$, where the small Kondo coupling ground state is antiferromagnetic). The Zeeman splitting is obtained and we are also able to reproduce the folding of the Brillouin zone in the N\'eel antiferromagnetic state. 
For both magnetically ordered states, we find that the evolution of the Fermi surface is either absent or very smooth and gradual upon varying $x$. This result is in contrast with the coherence breakdown which is predicted in the Kondo phases. A possible explanation for this difference may be obtained by considering the Kondo lattice limit ($x=1$). In this case, Kondo spins contribute to the formation of a large Fermi surface for the Kondo coherent state, while they do not contribute for the magnetically ordered states. The breakdown of coherence which is depicted here in Fermi surfaces spectra for the Kondo phase is thus related with the contribution of Kondo ions to forming strongly correlated fermionic quasiparticles. This breakdown of coherence is different from the breakdown of Kondo effect that distinguishes Kondo phases from pure magnetically ordered phases. Indeed, Kondo effect is still present in the non-coherent dilute Kondo phase. In our calculations, we did not consider the possibility of mixed states where magnetic order might coexists with Kondo effect. For such states, we expect coherence breakdown signatures that might be concomitant with Fermi surface reconstructions resulting from magnetic order.

\subsection{Phase diagram: emergence of two transitions in the paramagnetic Kondo phases\label{Section:Phasediagram}}

\begin{figure}
	\centering
	\vspace{-10pt}
	\includegraphics[width=0.80\textwidth]{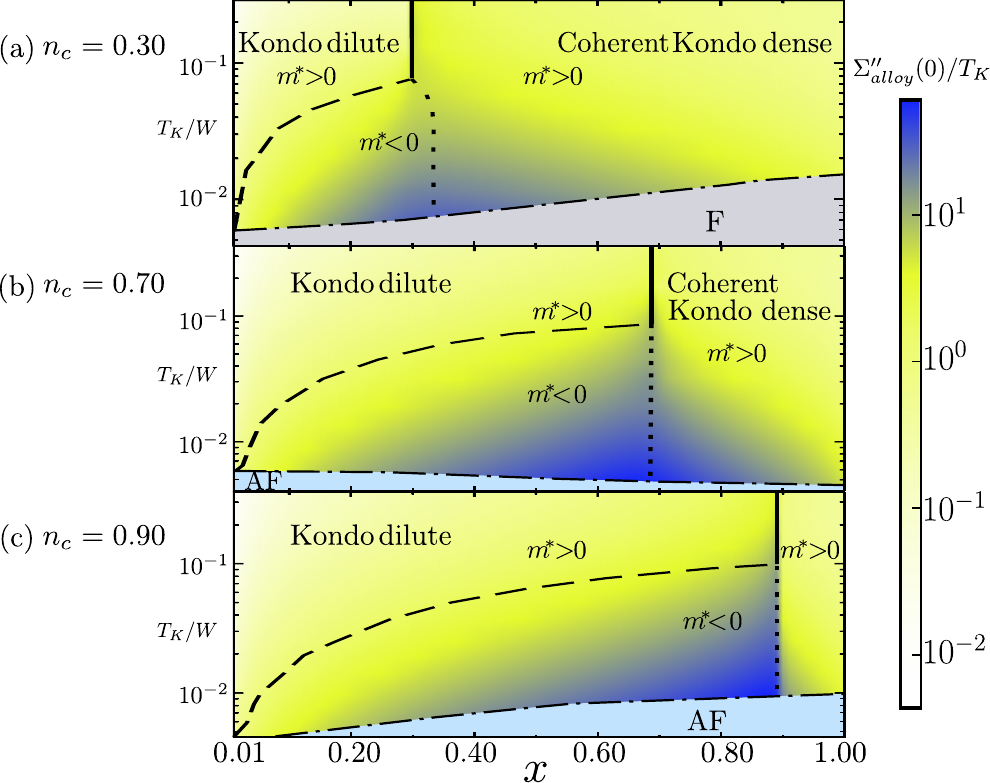}
	\caption{ \small{Ground state phase diagram of the KAM as functions of $x$ and $T_K/W$ for (a) $n_c = 0.30$, (b) $n_c = 0.70$ and (c) $n_c = 0.90$. In the Kondo phases the solid line indicates the discontinuity of the self-energy observed at $x=n_c$ for sufficiently strong $T_K$. This transition from the dense coherent Kondo phase becomes a crossover marked by an inflection in the self energy at smaller $T_K$ (doted line), and a significant increase in the intensity of imaginary part of the self-energy (color or black and white gradient). A continuous vanishing of the effective mass $m^\star$ is obtained at concentration $x^\star$ (dashed line), and we find $m^\star<0$ in the intermediate region
$x^\star < x< n_c$. We also solved the DMFT equations obtained for a Bethe lattice, considering the Kondo paramagnetic solution only and the same values of model parameters as depicted here. We were not able to distinguish the figures corresponding to the Bethe lattice from the ones depicted here for the 2D square lattice. This strong similarity excludes several interpretations that might invoke specificities of the lattice structure.
	\label{Figure:phasediagram}
	}}
\end{figure}

The phase diagram of the KAM is depicted in figure~\ref{Figure:phasediagram}. It was obtained by comparing the energies of each phase considered: paramagnetic Kondo, antiferromagnetic and ferromagnetic phases. From theses phase diagrams, we observe that in the region with small values of Kondo coupling, the long range magnetically ordered phases are stabilized. Their competitions and their stabilities for the KAM, which is consistent with Doniach argument, was discussed elsewhere~\cite{poudel2020phase}. For small to intermediate values of $T_K$, a paramagnetic Kondo phase with three distinct zones are identified: a dilute Kondo at $x \ll 1$, a dense Kondo at $x \approx 1$ separated by a large zone of intermediate state $x^\star < x < n_c$. When increasing the Kondo coupling, $x^\star$ tends towards $n_c$ and for strong coupling only dilute and dense Kondo phases are obtained, separated by a Lifshitz transition.

Now, we analyze the different regions of paramagnetic Kondo phases by the means of effective self-energy $\Sigma_{Alloy}(\omega)$ and its corresponding effective mass $m^\star$ with a focus on ARPES.
The transition observed in the Fermi surfaces (see section~\ref{section:elasticARPES}) at $x=n_c$ for sufficiently large values of Kondo coupling is characterized by a discontinuity of $\Sigma_{Alloy}(\omega=0)$ (solid line in figure ~\ref{Figure:phasediagram}). Since the real part of
the self-energy is related with a rescaling of the Fermi level, we interpret this transition as a signature of the Lifshitz transition
that was predicted elsewhere~\cite{burdin2013lifshitz} from a strong coupling approach of the Kondo alloy: for $x>n_c$, all magnetic degrees of freedom from conduction electrons are frozen by Kondo singlets formation. The Fermi liquid quasiparticles in this coherent dense regime are formed by the remaining degrees of freedom from unscreened Kondo impurities.
In the dilute regime $x<n_c$, the microscopic nature of quasiparticles is different and emerges from unscreened conduction electrons. Our result shows that this strong coupling picture may be realized for $T_K \gtrsim \frac{W}{10}$. Therefore, in order to observe a signature of coherence breakdown at $x=n_c$ from ARPES experiments in Kondo alloys, one would need to consider $f-$electron compounds with relatively large Kondo temperatures. In this case a valence fluctuation or valence transition might also become relevant as well but this issue is beyond the scope of the present work.

\begin{figure}
	\centering
	\includegraphics[width=\textwidth]{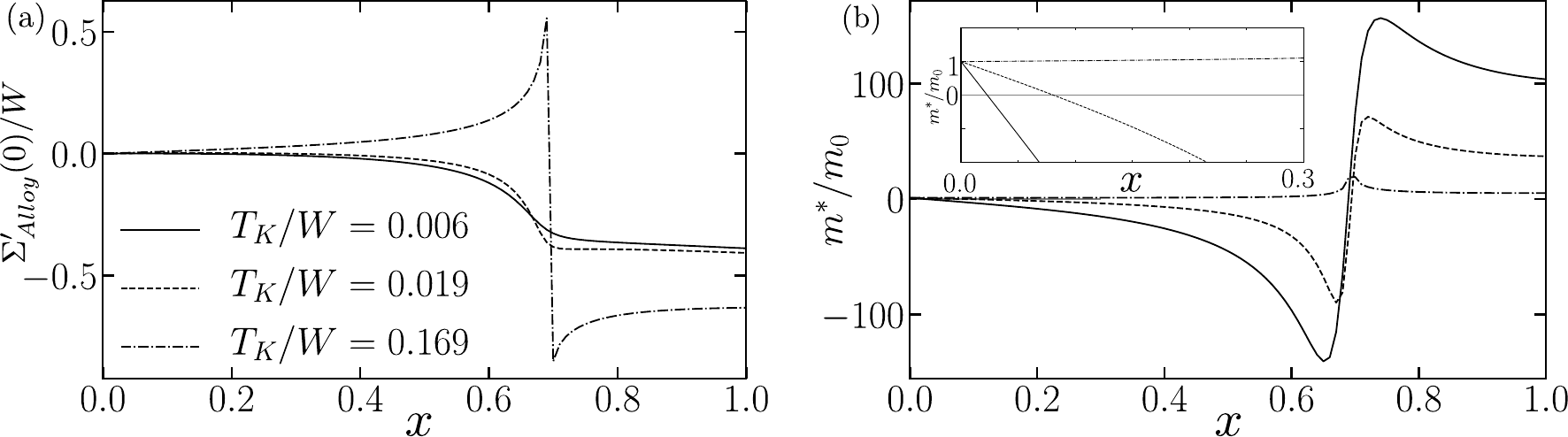}
	\caption{ \small{(a) Real part of the self-energy $\Sigma'_{Alloy}(0)/W$, (b) Effective mass $m^\star/m_0$
as functions of $x$, for $n_c = 0.70$.
	Different Kondo temperatures have been used for the numerics, illustrating the transition (at strong $T_K$) and crossover (at smaller $T_K$) obtained around $x=n_c$. The inset in (b) is a focus around the critical concentration $x^\star$ which is characterized by a vanishing of $m^\star$ when $T_K$ relatively small. At intermediate concentrations $x^\star<x<n_c$ we find $m^\star<0$.
	\label{Figure:sigmaprimedex}
	}}
\end{figure}

Hereafter, we analyze different situations with smaller Kondo coupling. The transition predicted at $x=n_c$ separating dense and dilute Kondo regimes becomes a crossover at smaller coupling and the quasiparticle lifetime is significantly shorten due to disorder incoherence effects. Indeed, the self-energy is found to be continuous and characterized by an inflection around this crossover $x=n_c$ (doted line in figure~\ref{Figure:phasediagram}, see also figure~\ref{Figure:sigmaprimedex}).
Photo-emission signatures of coherence breakdown in Kondo alloys: DMFT approach
Whereas, the imaginary part, $\Sigma''_{Alloy}(0)$, which is relatively small at strong coupling, becomes
large around this crossover at smaller coupling (see color/black-white gradient in figure~\ref{Figure:phasediagram}).

\begin{figure}
    \centering
	\includegraphics[width=0.55\textwidth]{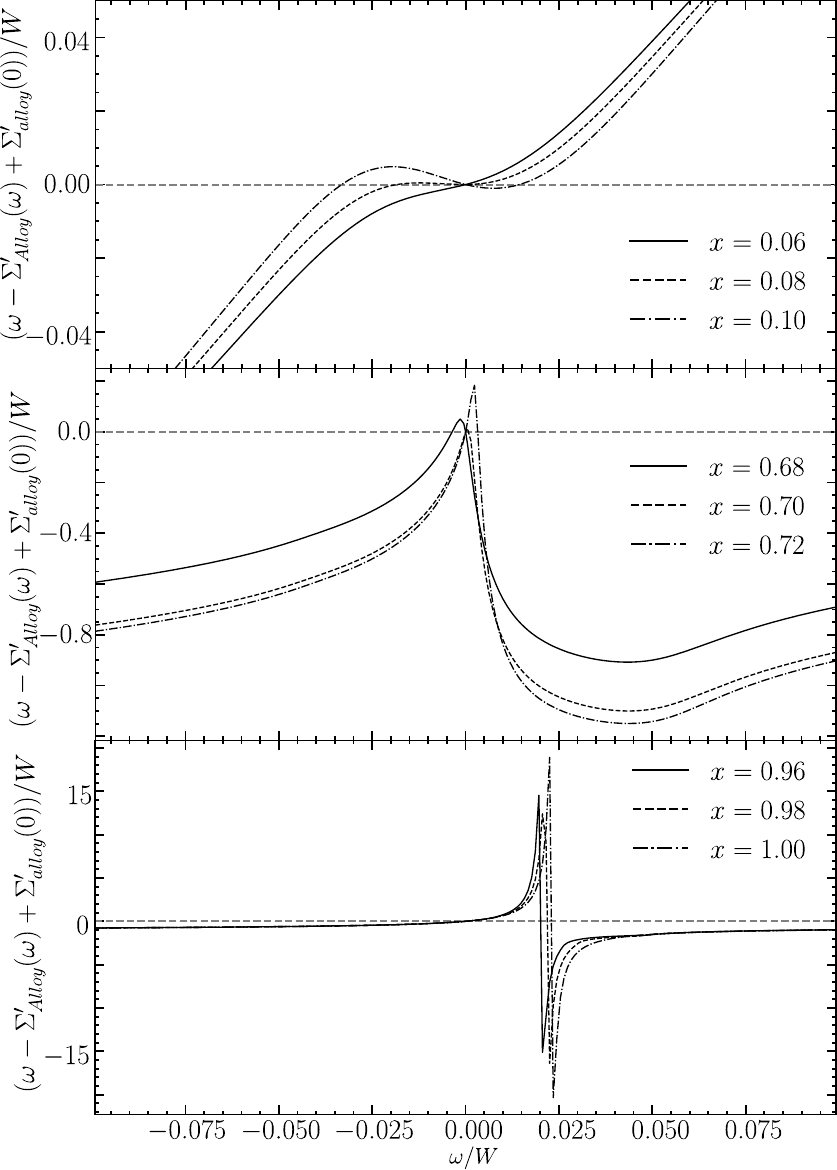}
	\caption{Frequency dependence of the real part of the self-energy, $\omega-\Sigma'_{Alloy}(\omega)+\Sigma'_{Alloy}(0$) for $n_c=0.70$ and $T_K=0.019$. Top: for $x$ in the vicinity of the critical point $x^\star=0.08$ which is characterized by $m^\star=0$, we observe the emmergence of a non-monotonicity at low energy. This leads to the gradual formation of a multiple-branches dispersion for $x>x^\star$.
	Center: for $x$ in the vicinity of $n_c$, the maximum is realized at $\omega<0$ for $x<n_c$ and at $\omega>0$ for $x>n_c$ resulting in a second change of sign of $m^\star$.
Bottom: for $x$ close to $1$, we observe signatures of the singularity $\Sigma_K(\omega)=\frac{r^2}{\omega+\lambda}$ which is obtained in the mean-field approximation for the Kondo lattice. The non-monotonicity obtained at lower concentrations is reminiscent of this singularity and we expect this feature to survive qualitatively beyond the mean-field approximation.
}
	\label{3_mass_effectives}
\end{figure}

In agreement with the ARPES spectral analyse of section~\ref{Section:inelasticARPES} for intermediate values of Kondo coupling, we found a third intermediate phase, which is separated from the dilute phase by a transition line in the $x-T_K$ phase diagram: at the critical concentration $x^\star$, the effective mass $m^\star$ vanishes continuously. This transition separates the very dilute regime ($x<x^\star$) with $m^\star>0$ from an intermediate regime ($x^\star<x<n_c$) characterized by $m^\star<0$ (see inset of figure~\ref{Figure:sigmaprimedex} and dashed line in figure~\ref{Figure:phasediagram}). From our numerical datas, we found a powerlaw relation $T_K/W \propto (x^\star)^\gamma$ with an exponent $\gamma < 1$ and the exponent itself depends on the filling $n_c$. We found $\gamma=$ 0.71, 0.78 and 0.83 for $n_c =$ 0.30, 0.70 and 0.90 respectively. This transition produces signatures in ARPES signal (see section~\ref{Section:inelasticARPES}) where two branches merge to one branch at $x^\star$ (see figure~\ref{Figure:inelasticARPESFaible}). Meanwhile, Fermi surface structures (see figure~\ref{Figure:ARPESElasticKondoPhase}) do not show a clear signature of $x^\star$. Keeping in mind that the dispersion relation is $\omega+\mu-\Sigma_{Alloy}(\omega)=\epsilon_{\bf k}$, the frequency dependence of the real part $\omega-\Sigma'_{Alloy}(\omega)+\Sigma'_{Alloy}(0)$ is represented in figure~\ref{3_mass_effectives}.
We can thus interpret the transition at $x=x^\star$ as the gradual formation (or extinction) of extra branches in the one-electron excitations, as suggested by the low frequency dependence of this quantity, which changes from locally monotonous (at $x<x^\star$) to locally non-monotonous (at $x>x^\star$). The second branch could then be associated with coherent and dispersive singlet-triplet excitations that may propagate. Such triplet excitations are gaped out at strong Kondo coupling but they might also reveal a pseudogap in PES at smaller Kondo coupling as analyzed in the section~\ref{Section:inelasticARPES}. This intermediate state is thus precursor to the coherent state which is realized around $x \approx 1$. It is very interesting to see that such a pre-coherent state may start being formed at a relatively small concentration $x^\star$.

Two very important features could make the transition at $x^\star$ observable experimentally: first, the imaginary part of
$\Sigma''_{Alloy}(0)$ remains relatively small around $x^\star$ (see color/black-white gradient in figure~\ref{Figure:phasediagram}). We can thus expect that the excitations are long lifetime quasiparticles which could be revealed by photo-emission. Secondly, $x^\star$ depends on the strength of the Kondo interaction. We may thus expect that this transition could be tuned by applying pressure on a compound with fixed concentration $x$. Of course, we are aware that mechanical pressure is not fully compatible with ARPES experiments. However, since the underlying phenomenon is a transition in the one-electron excitation spectrum, we may expect signatures in other sorts of experiments that could be realized under pressure, e.g., Raman spectroscopy.

\section{Conclusion} \label{section:conclusion}
In this theoretical paper we have investigated photo-emission spectra of disordered Kondo alloys. 
For this purpose, we have adapted the DMFT/CPA formalism that was developed elsewhere~\cite{georges1996dynamical,burdin2007random} to include the possibility of various ground states of the KAM: paramagnetic Kondo, antiferromagnetic and ferromagnetic phases. Using static mean-field theories as local impurity solvers for DMFT, we investigated a large range of values of electronic filling $n_c$, impurity concentrations $x$, and strength of Kondo coupling $T_K$. We studied both a square lattice and a Bethe lattice structure and we obtained consistent results when comparison was possible. For the sake of clarity, the ARPES signals are analyzed only for the square lattice, but we may expect that they can be generalized to other lattice structures. 
 
Our results confirm the existence of a transition at $x=n_c$ between a coherent dense Kondo regime  (for $x>n_c$) and a dilute Kondo regime (for $x<n_c$)~\cite{burdin2013lifshitz}. This Lifshitz-like transition may be observed in ARPES experiments through the analyse of Fermi surface for Kondo alloys with a Kondo temperature higher than about $1/10$ bandwidth. A shrinking of the Fermi-surface is expected when increasing $x$ in the dilute regime, while the Fermi surface is enlarge with $x$, in connection with Luttinger theorem, in the coherent dense Kondo regime. It could be obtained experimentally in the materials with a relatively large $T_K$. We have also shown that this transition at $x=n_c$ becomes a crossover at smaller values of Kondo interaction. 

While Doniach argument is recovered, with magnetically ordered ground states stabilized at very small Kondo coupling, we identified a broad region of parameters where an intermediate paramagnetic Kondo state can be stabilized. This corresponds to values of $T_K$ between $W/100$ and $W/10$ with the regime of concentrations $x^\star<x<n_c$ which can be realized in a large variety of heavy-fermion Kondo systems. Unlike the dilute and the dense Kondo states, this intermediate phase is characterized by a negative effective mass. Here, negative effective mass does not mean instability, but it corresponds to the emergence of an extra branch in the electronic dispersion and a formation of a pseudogap in the local density of states at $x>x^\star$. 

We also analyze the effects of decoherence resulting from disorder, which may spoil possibilities of analyzing ARPES signals. 
We find that these effects are relatively small at strong Kondo coupling, which may make possible the observability of the transition at $x=n_c$. However this requires compounds with a relatively large value of $T_K$.
At smaller values of Kondo interaction, disorder-induced decoherence may become significant, with a maximal effect around $x=n_c$. This disorder-induced decoherence reduces the opportunities of observing signatures in Fermi surface at $x=n_c$ for small coupling. We still expect relatively well defined quasiparticles around $x=x^\star$ even through contrary to the transition at $x=n_c$, the transition at $x=x^\star$ cannot be observed directly from the Fermi surface structures. Indeed, the experiments involving low energy excitations (e.g. ARPES) are more appropriate. This opens rich perspectives for experimental investigations of the 
breakdown of coherence in Kondo alloys. Since we predict that $x^\star$ varies with the strength of the Kondo interaction, this transition might be realized not only by atomic substitution but also by applying pressure in a Kondo alloy with fixed stoichiometry (i.e. fixed $x$ and $n_c$). On a fundamental point of view, this work also suggests the possible emergence of exceptional points~\cite{PhysRevB.98.035141, PhysRevB.101.085122} at $x^\star$ where two dispersives branches merge and $m^\star$ vanishes. The physical excitations of the quasiparticles should be investigated at $x=x^\star$ theoretically and experimentally by direct or indirect probes.

\ack
We acknowledge useful discussions with Andr\'es Santander-Syro, Christoph Geibel, Clemens Laubschat, Cornelius Krellner, Denis Vyalikh, Ilya Sheikin, Jochen Wosnitza, J\'er\'emy Sourd, Marcelo Rozenberg, Marie-Aude Measson, Sophie Tenc\'e and V\'eronique Brouet. This work has been supported by Agence National de Recherche (ANR) and Deutsche Forschungsgemeinschaft (DFG) Grant Fermi-NESt (ANR-16-CE92-0018).

\section*{Appendix}

\appendix

\section{Mean-field approximations} \label{appendix:MF_approximations}

\subsection{Weiss mean-field approximation for ferromagnetic (F) and antiferromagnetic (AF) phases} \label{appendix:MF_approximation_Weiss}
Here we describe the mean-field impurity solver that we used for solving the local effective action on a Kondo site, which is given by Eq.~(\ref{actionloc:kondo}), assuming a magnetically ordered state. 
We restrict our study to pure ordered phases, and we ignore the possibility of mixed phases where 
Kondo effect and magnetic order might coexist. Furthermore, we assume that the magnetic order is either ferromagnetic (F) or staggered antiferromagnetic (AF). 
Invoking the $A\leftrightarrow B$ symmetries of these phases, the single site effective DMFT/CPA action can be solved assuming this site belongs to the sublattice $A$. 
Consequently, hereafter, in this section we consider only $\alpha=$A and we omit the $\alpha$ index. 

The Kondo interaction is decoupled using the standard Weiss mean-field approximation as $\mathbf{S}\cdot\mathbf{s} \approx  \langle S^z \rangle s^z +  S^z  \langle s^z \rangle - \langle S^z \rangle  \langle s^z \rangle$. 
The order parameters are 
the local magnetizations on a Kondo site belonging to the sublattice $A$:  
$m_f\equiv\langle S^z \rangle$, and 
$m_c\equiv\langle s^z \rangle = 
\frac{1}{2} \sum_\sigma \sigma_z \langle c_{ \sigma}^\dagger c_{ \sigma} \rangle$. 
The local effective action~(\ref{actionloc:kondo}) becomes quadratic and the local Green function on a 
$\mathcal{K}$-site can be expressed explicitely in terms of the bath and order parameters as 
$G_{\mathcal{K} \sigma} (i\omega ) = 
\frac{1}{i\omega + \mu - \Delta_{\sigma}(i\omega) - \sigma J_K \frac{m_f}{2}}$. 

Finally, to complete the general DMFT/CPA self-consistent equations described in section~\ref{section:DMFTmethod} which relate the local Green functions and the dynamical bath, the chemical potential $\mu$ and the order parameters $m_f$ and $m_c$ are determined by solving the 
following self-consistent equations: 
\begin{eqnarray}
n_c &=& x\frac{1}{\beta} \sum_{i\omega,\sigma} G_{\mathcal{K} \sigma} (i\omega)+ (1-x)\frac{1}{\beta} \sum_{i\omega,\sigma}G_{\mathcal{N} \sigma} (i\omega)~, \label{ferro:MFequation1}\\
m_c &=&  \frac{1}{2\beta} \sum_{i\omega, \sigma} \sigma G_{\mathcal{K} \sigma} (i\omega)~, 
\label{ferro:MFequation2} \\  
m_f &=& -\frac{1}{2} \tanh(\frac{\beta m_c J_K}{2})~. \label{ferro:MFequation3}
\end{eqnarray}

\subsection{``Slave boson'', ``large-n'', mean-field approximation for the Kondo (K) phase} \label{appendix:MF_approximation_Kondo}
Here we describe the mean-field impurity solver that we used for solving the local effective action on a Kondo site, which is given by Eq.~(\ref{actionloc:kondo}), assuming a paramagnetic Kondo correlated state. We neglect the possibility that magnetic order might co-exist with Kondo local strong correlations. Consequently, hereafter, in this section we omit the 
$\alpha$ sublattice index. 
We follow the standard mean-field approximation as introduced by Lacroix and Cyrot in (~\cite{lacroix1979phase}), which is analogous to the ``large N" expansion or slave boson 
approximation developed by Coleman in (~\cite{coleman19831}) and Read and Newns in (~\cite{read1984stability}). 

At first, we represent Kondo spin operator within Abrikosov's fermionic representation 
$\mathbf{S}^{\sigma \sigma^\prime} = f_{\sigma}^\dagger f_{\sigma^\prime} - \delta_{\sigma \sigma^\prime}/2$.  Thereafter the 
Kondo interaction in Eq.~(\ref{actionloc:kondo}) is mapped as 
${\bf S}\cdot{\bf s}\to \frac{1}{2}\sum_{\sigma \sigma^\prime} c_{\mathcal{K} \sigma}^\dagger c_{\mathcal{K}\sigma^\prime} f_{\sigma^\prime}^\dagger f_\sigma$. 
Finally, the Kondo interaction is approximated invoking the mean-field decoupling 
$c_{\mathcal{K} \sigma}^\dagger c_{\mathcal{K}\sigma^\prime} f_{\sigma^\prime}^\dagger f_\sigma 
\approx  \langle c_{\mathcal{K}\sigma}^\dagger f_{\sigma}\rangle f_{\sigma^\prime}^\dagger c_{\mathcal{K}\sigma^\prime} 
+  \langle f_{\sigma^\prime}^\dagger c_{\mathcal{K}\sigma^\prime} \rangle c_{\mathcal{K}\sigma}^\dagger f_{\sigma} - \langle f_{\sigma^\prime}^\dagger c_{\mathcal{K}\sigma^\prime} \rangle  \langle c_{\mathcal{K} \sigma}^\dagger f_{\sigma}\rangle $. 
This mean-field description results in an emergent effective hybridization between the conduction electrons and the Abrikosov fermions, 
$r= \frac{J_K}{2} \sum_{\sigma} \langle c^\dagger_{\mathcal{K}\sigma} f_{\sigma} \rangle $, which can be identified 
as an order parameter for the Kondo phase. 
An additional constraint $\sum_{\sigma} f_{\sigma}^\dagger f_{\sigma} = 1 $  restricts the number of Abrikosov 
fermions to one, which are imposed by introducing a Lagrange parameter $\lambda (\tau)$. 
With the mean-field approximation, $\lambda$ assumed to be constant and determined self-consistently in order to satisfy the $f$ occupancy constraint on average. 

Since here we are considering paramagnetic Kondo phase, for the sake of clarity, hereafter we skip the spin index $\sigma$. 
The local effective action~(\ref{actionloc:kondo}) becomes quadratic and the local Green function on a 
$\mathcal{K}$-site can be expressed explicitly in terms of the bath and order parameters as 
$G_{\mathcal{K}}(i\omega)  = 
\frac{1}{i\omega + \mu - \Delta_{\sigma}(i\omega) - \frac{r^2}{i\omega+\lambda}}$. 
The Kondo self-energy involved in Eq.~(\ref{equation:kondoselfenergy}) thus has a pole singularity 
$\Sigma_K(i\omega)=\frac{r^2}{i\omega+\lambda}$ which captures several relevant aspects of Kondo physics, including the two-branches ARPES signature of the Kondo lattice ($x=1$) coherent state. 

Finally, to complete the general DMFT/CPA self-consistent equations described in section~\ref{section:DMFTmethod} which relate the local Green functions and the dynamical bath, the chemical potential $\mu$, the order parameter $r$, 
and the Lagrange multiplier $\lambda$ are determined by solving the 
following self-consistent equations: 
\begin{eqnarray}
n_c &=& x\frac{2}{\beta} \sum_{i\omega} G_{\mathcal{K}} (i\omega)+ (1-x)\frac{2}{\beta} \sum_{i\omega}G_{\mathcal{N}} (i\omega)~, \label{Kondo:MFequation1}\\
\frac{r}{J_K}
 &=&  -\frac{r}{\beta} \sum_{i\omega} \frac{G_{\mathcal{K}} (i\omega)}{i\omega+\lambda}~, 
\label{Kondo:MFequation2} \\  
1 &=&  \frac{2}{\beta} \sum_{i\omega} \frac{r^2G_{\mathcal{K}} (i\omega)}{(i\omega+\lambda)^2}~. \label{Kondo:MFequation3}
\end{eqnarray}

\section*{References}
\bibliographystyle{unsrt}
\bibliography{biblioKondo_mstarPES}

\end{document}